\documentclass[10pt, conference, compsocconf]{IEEEtran}

\usepackage{graphicx}
\begin{document}

\title{Simulation Analysis of\\ Medium Access Techniques}

\author{I. Israr, M. M. Yaqoob, N. Javaid, U. Qasim$^{\ddag}$, Z. A. Khan$^{\S}$\\

        $^{\ddag}$University of Alberta, Alberta, Canada\\
        Department of Electrical Engineering, COMSATS\\ Institute of
        Information Technology, Islamabad, Pakistan. \\
        $^{\S}$Faculty of Engineering, Dalhousie University, Halifax, Canada.
        }

\maketitle

\begin{abstract}
This paper presents comparison of Access Techniques used in Medium Access Control (MAC) protocol for Wireless Body Area Networks (WBANs). Comparison is performed between Time Division Multiple Access (TDMA), Frequency Division Multiple Access (FDMA), Carrier Sense Multiple Access with Collision Avoidance (CSMA/CA), Pure ALOHA and Slotted ALOHA (S-ALOHA). Performance metrics used for comparison are throughput (T), delay (D) and offered load (G). The main goal for comparison is to show which technique gives highest Throughput and lowest Delay with increase in Load. Energy efficiency is major issue in WBAN that is why there is need to know which technique performs best for energy conservation and also gives minimum delay.
\end{abstract}

\begin{IEEEkeywords}
Pure ALOHA, Slotted ALOHA, CSMA/CA, TDMA, FDMA, Wireless Body Area Networks, Throughput, Delay, Offered Load
\end{IEEEkeywords}

\section{Introduction}

Energy efficiency is an important issue in WBANs, because sensor nodes damage human body tissue. More importantly sensor nodes connected to body are battery operated devices, they have limited life time. So, MAC protocols of WBANs need to be energy efficient and supports medical applications. It allows integration of low power intelligent sensor nodes. They are used to stream biological information from human body and transmit it to a coordinator. This procedure is very helpful while monitoring health of a person and in case of emergency providing proper medication. MAC protocol plays an important role in determining the energy efficiency of a protocol in WBANs. Traditional MAC protocols focus on improving throughput and bandwidth efficiency. However, the most important thing is that they lack in energy conserving mechanisms. The main source of energy wastage are idle listening, overhearing and packet overhead. Controlling these energy waste sources maximizes network lifetime.
\\
\indent WBANs have many advantages like mobility of patient and independent monitoring of patient. It can work on Wireless Local Area Networks (WLANs), Worldwide Interoperability for Microwave Access (WiMAX) or internet to reliably transmit data to a server which is monitoring health issues. There are some requirements for the MAC protocol design to be used in WBANs. Firstly all of protocols must have high QoS (Quality of Service), it must be reliable, it needs to support different medical applications.
\\
\indent By using different Medium Access Techniques, different low power and energy efficient protocols for MAC are proposed. The most important attributes of WBANs are low power consumption and delay. Different techniques are used with different protocol to control the delay and to improve the efficiency of MAC protocol. Techniques like Energy Efficient low duty cycle MAC protocol [1], Traffic Adaptive MAC protocol [3], Energy Efficient TDMA based MAC protocol [4] are used to improve energy efficiency and to control delay.
\\
\indent The important techniques of MAC protocol for WBANs are Time Division Multiple Access (TDMA) and Carrier Sense Multiple Access with Collision Avoidance (CSMA/CA). Frequency Division Multiple Access (FDMA) is very close to TDMA. Pure Aloha and SLOTTED Aloha are not used due to collision problems and high packet drop rates as well as low energy efficiency. There are several challenges in realization of the perfect Multiple Access Technique for MAC protocol design.
\section{Related Work And Motivation}
Authors in [1] state that the IEEE 802.15.4 standard is designed as a low power and low data rate protocol with high reliability. They analyze unslotted version of protocol with maximum throughput and minimum delay. The main purpose of 802.15.4 is to give low power, low cost and reliability. This standard defines a physical layer and a MAC sub layer. It operates in either beacon enabled or non beacon mode. Physical layer specifies three different frequency ranges: 2.4 GHz band with 16 channels, 915 MHz with 10 channels and 868 MHz with 1 channel. Calculations are performed by considering only beacon enabled mode and with only one sender and receiver. However, it is high power consumed standard. As number of sender increases, efficiency of 802.15.4 decreases. Throughput of 802.15.4 declines and delay increases when multiple radios are used because of increase in number of collisions.\\
\indent Energy Efficient TDMA based MAC Protocol is described in [2]. Protocol in this paper minimizes the amount of idle listening by sleep mode this is to reduce extra cost for synchronization. It listens for synchronization messages after a number of time frames which results in extremely low communication power. However, this protocol lacks wake-up radio mechanism for on demand traffic and emergency traffic.\\
\indent In paper [3], authors propose using a wake-up radio mechanism MAC protocol for wireless body area network. Comparison of TDMA with CSMA/CA is also done in this paper. Proposed MAC protocol save energy by node going to sleep when there is no data and can be waked up on-demand by wake-up radio mechanism. This protocol works on principle of on-demand data. It reduces the idle time consumption of a node to a great extent. However, emergency traffic are not discussed in this paper, which is a major issue in WBANs.\\
\indent An Ultra Low Power and Traffic adaptive protocol designed for WBANs is discussed in [4]. They used a traffic adaptive mechanism to accommodate on-demand and emergency traffic through wake-up radio. Wake-up radio is low power consumption technique because it uses separate control channel with data channel. Comparison of power consumption and delay of TA-MAC with IEEE 802.15.4, Wise MAC, SMAC are done in this paper.\\
\indent Authors describe energy efficient low duty cycle MAC protocol for WBANs in paper [5]. TDMA are compared with CSMA/CA. TDMA based protocol outperforms CSMA/CA in all areas. Collision free transfer, robustness to communication errors, energy efficiency and real time patient monitoring are the flaws that are overcome in this paper. However, synchronization is required while using TDMA technique. With increase in data, TDMA energy efficiency decreases due to queuing. As network topology changes TDMA experiences degradation in performance.\\
\indent In this paper [6], authors introduce a context aware MAC protocol which switch between normal state and emergency state resulting in dynamic change in data rate and duty cycle of sensor node to meet the requirement of latency and traffic loads. Also they use TDMA frame structure to save power consumption. Additionally a novel optional synchronization scheme is propose to decrease the overhead caused by traditional TDMA synchronization scheme. However, throughput in this paper is not addressed.\\
\indent In paper [7], authors propose technique for mechanism of low power for WBAN, that defines traffic patterns of sensor nodes to ensure power efficient and reliable communication. They classify traffic pattern into three different traffic patterns (normal traffic, on-demand traffic and emergency traffic) for both on-body and in-body sensor networks. However they have not taken care for the delay and throughput. Also complete implementation of their proposed protocol is still to be done.\\
\indent PHY and MAC layers of IEEE 802.15.6 standard are discussed by author in paper [8]. They stated specifications and identified key aspects in both layers. Moreover bandwidth efficiency with increase in payload size is also analyzed. They also discuss the different modes of security in the standard. However, bandwidth efficiency of the standard is only investigated for CSMA/CA. Also they not discussed throughput and delay.\\
\indent In paper [9], author proposes a modified MAC protocol for WBAN which focuses on simplicity, dependability and power efficiency. It is used in Contention Access period and CSMA/CA is used in contention free period. Data is transmitted in the contention free period where as CAP is only used for Command packets and best effort data packets. However, propagation delay is not neglected which we consider in our comparison and also interference from other WBAN nodes are not taken into account while doing calculation. Technique which are used by the author have high delay as compared to TDMA and FDMA.\\
\indent Authors evaluate performance of IEEE 802.15.4 MAC, Wise MAC, and SMAC protocols for a non-invasive WBAN in terms of energy consumption and delay in [10]. IEEE 802.15.4 MAC protocol are improved for low-rate applications by controlling the beacon rate. In addition, beacons are sent according to the wakeup table maintained by the coordinator. However, authors have not discussed delay and offered load in their paper.\\
\indent In this [11], authors propose a new protocol MedMAC and they elaborate novel synchronization mechanism, which facilitates contention free TDMA channels, without a prohibitive synchronization overhead. They focus on power efficiency of MedMAC. Also they show that MedMAC performs better than IEEE 802.15.4 for very low data rate applications, like pulse and temperature sensors (less than 20 bps). However, they have discussed about collisions but they have not focused on delay in the applications.\\
\indent Authors in [12] presents implementation of energy efficient real time on demand MAC protocol for medical Wireless Body Sensor Network. They introduced secondary channel for slave sensor node for channel listening in idle state. This secondary channel brings in the benefits of acquiring zero-power from slave sensor node battery when listening, to achieve very-low-power. Nevertheless problem arises because of the tradeoff between low-power and real time wake up. However, they have not discussed about time critical application which require higher throughput and priority.\\
\indent In this paper [13], authors introduced a TDMA-based energy efficient MAC protocol for in-vivo communications between mobile nodes in BSNs using uplink/downlink asymmetric network architecture. They also proposed TDMA scheduling scheme and changeable frame formats. The latency optimization is discussed and the performance is improved by reducing the data slot duration. However they have not elaborated about throughput and delay sensitive application.\\

\section{Introduction Of Multiple Access Techniques}
Channel access mechanisms provided by Medium Access Control (MAC) layer are also expressed as multiple access techniques. This made it possible for several stations connected to the same
physical medium to share it. Multiple access techniques have been used in different type of networks. Each technique is used according to its requirement.
In this paper, we are comparing behavior of different multiple access techniques with change in throughput, delay and offered load.
We have discuss them considering three scenarios.
\\
{(1)} Offered load as a function of delay.
\\
{(2)} Offered load as a function of throughput.
\\
{(3)} Throughput as a function of delay.

\subsection{TDMA}
TDMA works with principle of dividing time frame in dedicated time slots, each node sends data in rapid
succession one after the other in its own time slot. Synchronization is one of the key factors while applying TDMA. It uses full channel width, dividing it
into two alternating time slots. TDMA uses less energy than others due to less collision and no idle listening. TDMA protocols are more power efficient than other multiple access protocols because nodes transmits only in allocated time slots and all the other time in inactive state. A packet generated by node suffer three type of delays as it reaches receiver.
\\
{(1)}Transmission delay.
\\
{(2)}Queuing delay.
\\
{(3)}Propagation delay.
\\Equations which we have used to plot TDMA in three scenarios are given below:
\begin{figure}[!h]
\centering
\caption{Timing diagram of TDMA}
\includegraphics[width=3 in, height=2 in]{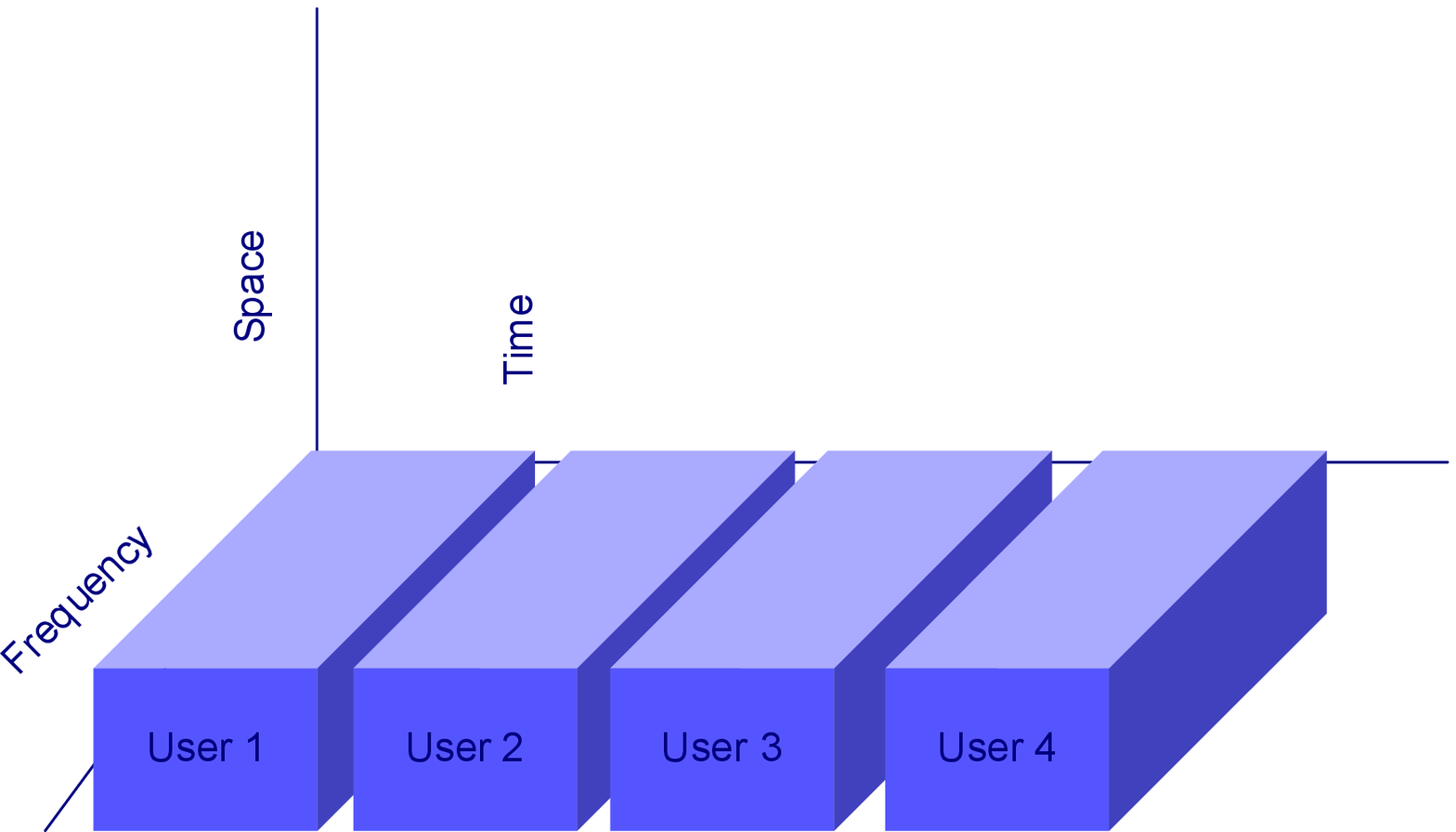}
\end{figure}
\\
Relation of D and T
\begin{eqnarray}
D=L/C+q/(2*(1-q))*N*L/C+N*L/(2*C)
\end{eqnarray}
\\
Relation of G and T
\begin{eqnarray}
T=L/C+G/(2*(1-G))*N*L/C*N*L/(2*C)
\end{eqnarray}
\\
Relation of D and G
\begin{eqnarray}
D=L/a+q./(2*(1-q))*N*L/a+N*L/(2*a)
\end{eqnarray}
\begin{table}
\caption{Description of Parameters used in Equations}
\begin{center}
    \begin{tabular}{ | p{2.5cm} | p{2.5cm} |}
    \hline
    Parameters & Description\\ \hline
    D & Delay\\ \hline
    T & Throughput\\ \hline
    C & Cycle length\\ \hline
    L & Number of packets\\ \hline
    N & Number of nodes\\ \hline
    q & Number of packets in queue\\ \hline
    G & Offered load\\ \hline
    a & Normalized end-to-end delay\\ \hline
    K & Kaapa\\ \hline
    \end{tabular}
\end{center}
\end{table}
\subsection{FDMA}
FDMA is a basic technology in analog Advanced Mobile Phone Service (AMPS), most widely-installed cellular phone system installed in North America. With FDMA, each channel can be assigned to only one user at a time.
Each node share medium simultaneously though transmits at single frequency. FDMA is used with both analog and digital signals. It requires high-performing filters in radio hardware, in contrast to TDMA and CSMA. As each node is separated by its frequency, minimization of interference between nodes is done by sharp filters. In FDMA a full frame of frequency band is available for communication, In FDMA a continuous flow of data is used, which improves efficiency of sending data. The division of frequency bands among users is shown in Fig 2.
\begin{figure}[!h]
\centering
\caption{Frequency distribution in FDMA}
\includegraphics[width=3 in, height=2 in]{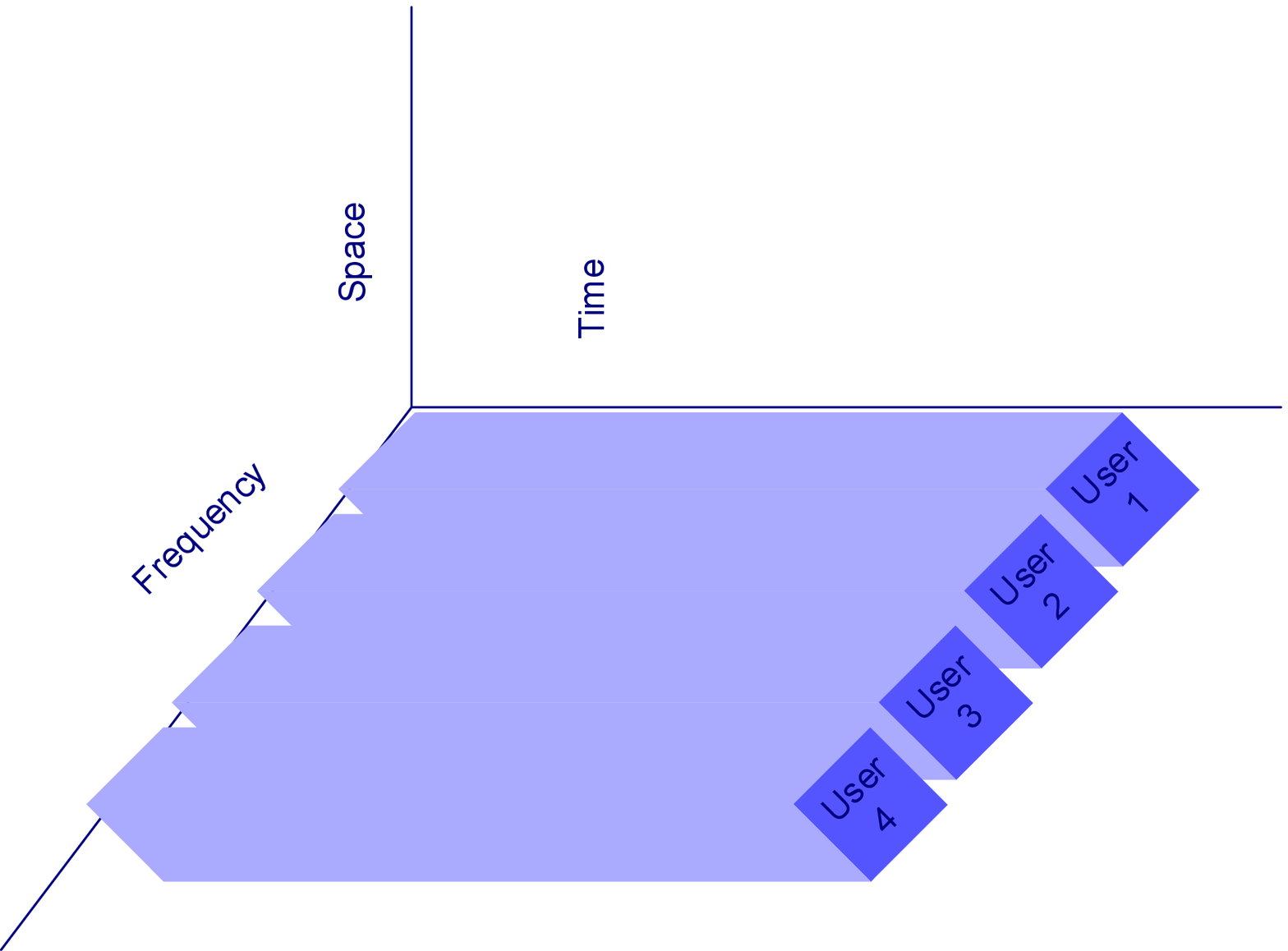}
\end{figure}
\\Relation of D and G
\begin{eqnarray}
D=N*L/a+q/(2*(1-q))*N*L/a
\end{eqnarray}
Relation of D and T
\begin{eqnarray}
D=N*L/C+q/(2*(1-q))*N*L/C
\end{eqnarray}
Relation of T and G
\begin{eqnarray}
T=N*L/C+G/(2*(1-G))*N*L/C
\end{eqnarray}
\subsection{CSMA/CA}
CSMA/CA is a extended version of CSMA. Collision avoidance is used to enhance performance of CSMA by not allowing node to send data if other nodes are transmitting. In normal CSMA nodes sense the medium if they find it free, then they transmits the packet without noticing that another node is already sending the packet, this results in collision. To improve the probability of collision CSMA/CA was proposed, CSMA/CA results in the improvement of collision probability.\\
\indent It works with principle of node sensing medium, if it finds medium to be free, then it sends packet to receiver. If medium is busy then node goes to back-off time slot for a random period of time and wait for medium to get free. With improve CSMA/CA RTS/CTS exchange technique node send Request to send (RTS) to receiver after sensing the medium and finding it free. After sending RTS, node waits
for Clear To Send (CTS) message from receiver. After message is received, it starts transmission of data, if node does not receive CTS message
then it goes to back-off time and wait for medium to get free. CSMA/CA is a layer 2 access method.  It is used in 802.11 wireless LAN and other wireless communication.
\begin{figure}[!h]
\centering
\caption{Timing diagram of CSMA/CA}
\includegraphics[width=3.5 in, height=2 in]{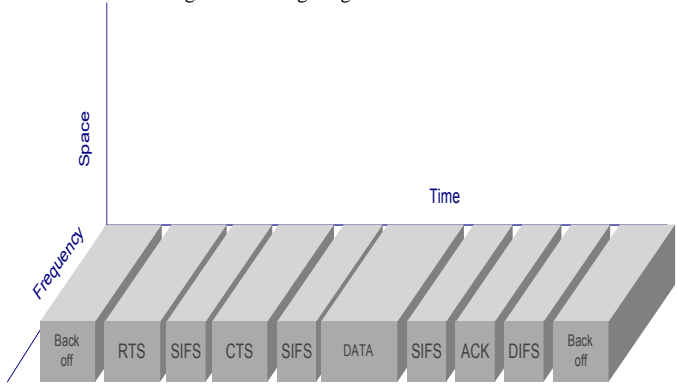}
\end{figure}
\\
Equations which we have used for plotting of CSMA/CA in three scenarios are given below
\\
Relation of D and T
\begin{eqnarray}
D=(exp(2*T)-1)*((K-\nonumber\\
1)/(2+2*a+1)+1+a.
\end{eqnarray}
\\
Relation of D and G
\begin{eqnarray}
D=(G.*(1+G+L.*G.*(1+G+L.*G./2)).\nonumber\\
*exp-G.(*(1+2*L)))./(G.*(1+2*L)-\nonumber\\
(1-exp(-L.*G))+(1+L.*G).*exp(-G \nonumber\\
.*(1+L)))
\end{eqnarray}
\\
Relation of G and T
\begin{eqnarray}
T=(G.*(1+G+a.*G.*(1+G+a.*G./2 \nonumber\\
)).*exp(-G.*(1+2*a)))./(G.*(1+2  \nonumber\\
*a)-(1-exp(-a.*G))+(1+a.*G).   \nonumber\\
*exp(-G.*(1+a)))
\end{eqnarray}
\\
\subsection{Pure ALOHA}
Pure ALOHA is the first random access technique introduced and it is so simple that its implementation is straight forward. It belongs to the family of contention-based protocols, which do not guarantee the successful transmission in advance. In this whenever a packet is generated, it is transmitted immediately without any further delay. Successful reception of a packet depends only whether it is collided or not with other packets. In case of collision, the collided packets are not received properly. At the end of packet transmission each user knows either its transmission successful or not.
\\
\indent If collision occurs, user schedules its re-transmission to a random time. The randomness is to ensure that same packet do not collide repeatedly. An example of Pure ALOHA is depicted in Fig 6. Each packet is belongs to a separate user due to the fact that population is large.
\begin{figure}[!h]
\centering
\caption{Pure ALOHA timing diagram}
\includegraphics[width=3 in, height=3 in]{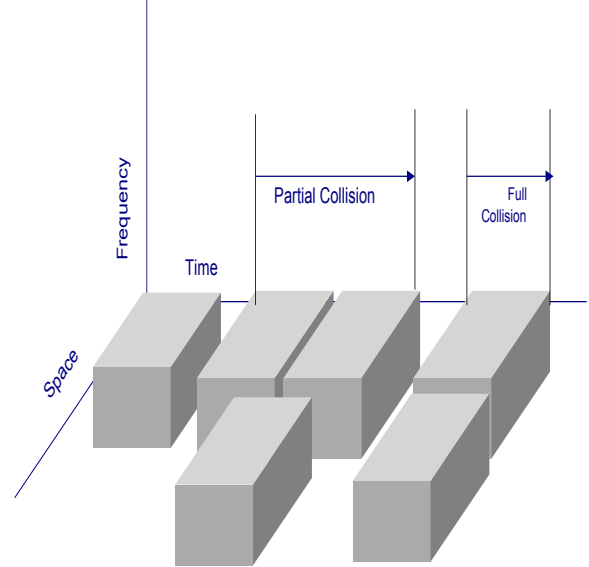}
\end{figure}
\\
Relation Of T and G
\begin{eqnarray}
T=&G*exp(-2*G)
\end{eqnarray}
Relation of T and D
\begin{eqnarray}
D=&(exp(2*S)-1)*((K- \nonumber\\
&1)/2+2*a+1)+1+a
\end{eqnarray}
Relation of D and G
\begin{eqnarray}
D=&(exp(G)-1*((K-1)/ \nonumber\\
&2+2*a+1)+1+a
\end{eqnarray}

\subsection{Slotted ALOHA}
Slotted ALOHA is a variant of Pure ALOHA with channel is divided into slots. Restriction is imposed on users to start transmission on slot boundaries only. Whenever packets collide, they overlap completely instead of partially. So only a fraction of slots in which packet is collided is scheduled for re-transmission. It almost doubles the efficiency of Slotted ALOHA as compared to Pure ALOHA.
Functionality of Slotted ALOHA is shown in Fig 5. Successful transmission depends on the condition that, only one packet is transmitted in each frame. If no packet is transmitted in a slot, then slot is idle. Slotted Aloha requires synchronization between nodes which lead to its disadvantage.
\begin{figure}[!h]
\centering
\caption{Slotted ALOHA timing diagram}
\includegraphics[width=3 in, height=3 in]{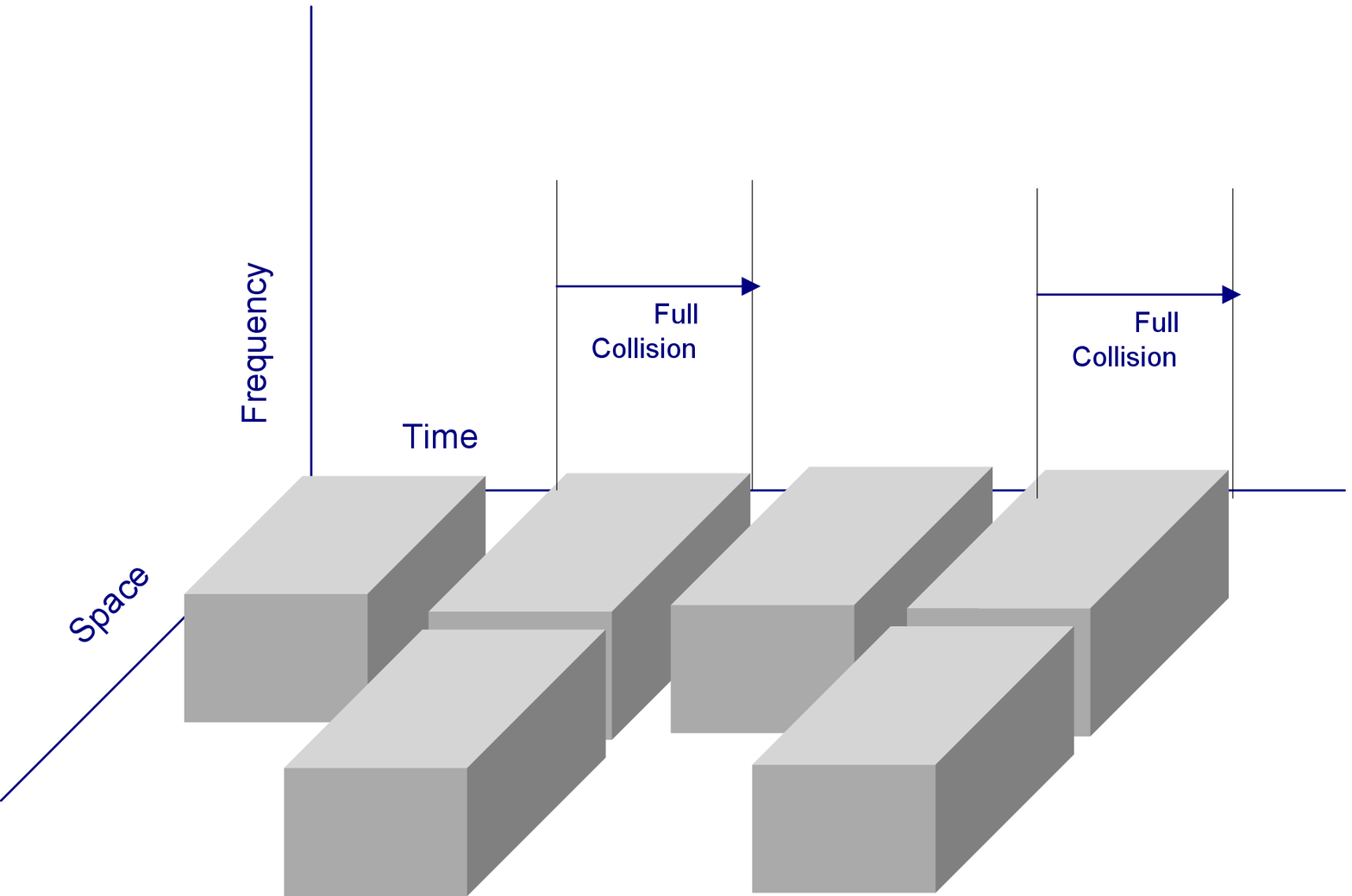}
\end{figure}
Relation of T and G
\begin{eqnarray}
T&=&G*exp(-G)
\end{eqnarray}
Relation of T and D
\begin{eqnarray}
D&=&(exp(S)-1)*((K-1)\nonumber\\
& &/2+2*a+1)+1.5+a
\end{eqnarray}
Relation of D and G
\begin{eqnarray}
D&=&(exp(G)-1)*((K-1)/ \nonumber\\
& &2+2*a+1)+1.5+a
\end{eqnarray}
\\
\subsection{Mathematical Modeling Of Throughput for Multiple Access Techniques}
In this section we are going to calculate the throughput of different multiple access techniques. Data is transferred from sender to receiver using one of the techniques, throughput due to these techniques have been calculated. Due to less difference between sender and receiver, there are no packet losses due to collision, no packets are lost due to buffer overflow. For the calculation of throughput we are assuming a perfect channel. Throughput is calculated for all access techniques through following equation.

\begin{eqnarray}
T&=& \frac{8.x}{delay(D) (x)}
\end{eqnarray}
In equation 16 D is delay, T is throughput and $x$ is the number of bits passing through the frame.
\subsubsection{Throughput of CSMA/CA}

Throughput of CSMA/CA is calculated by formula given in equation 16. Delay in equation 17 is calculated by adding delays of all elements of frame while it reaches receiver.

\begin{eqnarray}
D=T_{bo}+T_{data}+T_{ta}+T_{ack}+T_{ifs}+T_{rts}+T_{cts}
\end{eqnarray}
\\
The following notations are used: $T_{bo}=Back$ $Off$ $Period$, $T_{rts}=Request$ $To$ $Send$, $T_{cts}=Clear$ $To$ $Send$, $T_{data}=Transmission$ $Time$ $of$, $Data$, $T_{ta}=Turn$ $Around$ $Time$, $T_{ack}=Acknowledgement$ $Transmission$ $Time$, $T_{ifs}=Inter$ $Frame$ $Space$. \\
Now we calculate delay time given in equation 17
\begin{eqnarray}
T_{bo}=bo_{slots}.T_{boslots}
\\
T_{ta}=T_{data} + T_{ack}
\end{eqnarray}
\\
$bo_{slots}=Back$ $off$ $slots$ $number$
\\
$T_{boslots}=Back$ $off$ $slots$ $time$
\\
\begin{eqnarray}
T_{ack}&=&\frac{N_{ack}}{f_{c}}
\\
T_{ifs}&=&T_{data}-T_{ack}
\end{eqnarray}
Following notations are used: $T_{ta}=Turnaround$ $Time$, $T_{ack}=Acknowledgement$ $time$, $f_{c}= Communication$ $Data$ $Rate$, $N_{ack}=ACK/NACK$ $message$ , $bits$.
\indent If there is no acknowledgement then turnaround time $T_{turnaround}$ and $T_{ack}$ is equal to zero.
\\
\subsubsection{Throughput Of TDMA}

Throughput is calculated by using equation 16. Delay which a packet experiences as it reaches from sender to destination is calculated as following

\begin{eqnarray}
D=T_{oh}+T_{ack}+T_{g}+T_{sync}+T_{ta}
\end{eqnarray}
\\
Different time delay given in equation 22 can be calculated by following equations
\\
\begin{eqnarray}
T_{oh} & =&\frac{N_{oh}}{f_{c}}
\\
T_{ack}&=&\frac{N_{ack}}{f_{c}}
\\
T_{sync}&=&\frac{N_{syn}}{f_{c}}
\\
T_{data}&=&\frac{N_{data}}{f_{c}}
\end{eqnarray}

Following notations are used: $T_{sync}=Synchronization$ $time$, $T_{data}=Time$ $for$ $data$ $to$ $reach$ $end$ $of$ $frame$, $T_{ta}=Turnaround$ $Time$, $T_{ack}=Acknowledgement$ $time$, $T_{oh}=OverHead$ $time$, $T_{g}=Guard$ $time$, $f_{c}= Communication$ $Data$ $Rate$, $N_{oh}=Total$ $overhead$ $bits$, $N_{ack}=ACK/NACK$ $message$ $bits$, $N_{syn}=Total$ $synchronized$ $bits$, $N_{data}=Total$ $data$ $bits$.
\subsubsection{Throughput of FDMA}
Throughput of FDMA is very close to TDMA. There is very little difference between throughput of the two multiple access techniques. The calculation for the throughput of FDMA is calculated by formula given in equation 16 and the delay which it experience is calculated below
\begin{eqnarray}
D=T_{oh}+T_{ack}+T_{g}+T_{ta}+T_{data}
\end{eqnarray}
\\
Different time delay given in equation 27 can be calculated by following equations
\\
\begin{eqnarray}
T_{oh}&=&\frac{N_{oh}}{f_{c}}
\\
T_{ack}&=&\frac{N_{ack}}{f_{c}}
\\
T_{data}&=&\frac{N_{data}}{f_{c}}
\end{eqnarray}

Following notations are used
\\
$T_{data}=Time$ $for$ $data$ $to$ $reach$ $end$ $of$ $frame$
\\
$T_{ta}=Turnaround$ $Time$
\\
$T_{ack}=Acknowledgement$ $time$
\\
$T_{oh}=OverHead$ $time$
\\
$T_{g}=Guard$ $time$
\\
$f_{c}= Communication$ $Data$ $Rate$
\\
$N_{oh}=Total$ $overhead$ $bits$
\\
$N_{ack}=ACK/NACK$ $message$ $bits$
\\
$N_{data}=Total$ $data$ $bits$
\\
\subsubsection{Throughput of Pure ALOHA}
The calculation for the throughput of ALOHA is done by formula given in equation 16 and the delay which it experience is calculated below
\begin{eqnarray}
D=T_{data} + T_{que}
\end{eqnarray}
\\
Following notations are used: $T_{data}=Time$ $for$ $data$ $to$ $reach$ $end$ $of$ $frame$
\\
$T_{que}=Time$ $for$ $queuing$
\\
\subsubsection{Throughput of S-ALOHA}
The calculation for the throughput of S-ALOHA is done by formula given in equation 16 and the delay which it experience is calculated below
\begin{eqnarray}
D=T_{ack}+T_{syn}+T_{ta}+T_{idle}+ T_{bon}
\end{eqnarray}
\\
Different Time delay given in equation 32 can be calculated by following equations
\begin{eqnarray}
T_{ack}&=&\frac{N_{ack}}{f_{c}}
\\
T_{sync}&=&\frac{N_{sync}}{f_{c}}
\end{eqnarray}

Following notations are used: $T_{bon}=Time$ $for$ $data$ $to$ $be$ $transmitted$ $at$ $slot$ $boundaries$, $T_{idle}=Idle$ $time$ $after$ $a$ $transmission$, $T_{ta}=Turnaround$ $Time$, $T_{ack}=Acknowledgement$ $time$, $N_{syn}=Total$ $synchronized$ $bits$, $f_{c}= Communication$ $Data$ $Rate$, $N_{ack}=ACK/NACK$ $message$ $bits$.
\section{Conclusion}
In this paper different Multiple Access Techniques of MAC protocol which are used in Wireless Body Area Networks have been compared. Techniques are
TDMA, FDMA, CSMA/CA, ALOHA and SALOHA. Algoritham for all these techniques are given in this paper showing their working. Mathematical equations for the calculation of throughput for all these technqiues have been shown. Performance metrices for the comparison of these techniques are Throughput, Delay and Offered Load. Comparison has been done between performance metirces Throughput and Delay, Delay and Offered Load and Offered Load and Throughput. TDMA is the best technique used in WBAN with increase in load because it has highest throughput and minimum delay which is the most important requirement of Wireless Body Area Networks.

\end{document}